\documentclass[aps,prd,preprint,superscriptaddress]{revtex4}

\usepackage{graphicx}

\def\bwt{\begin{widetext}}
\def\ewt{\end{widetext}}
\def\be{\begin{equation}}
\def\ee{\end{equation}}
\def\bea{\begin{eqnarray}}
\def\eea{\end{eqnarray}}
\def\bean{\begin{eqnarray*}}
\def\eean{\end{eqnarray*}}
\def\bary{\begin{array}}
\def\eary{\end{array}}
\def\bit{\begin{itemize}}
\def\eit{\end{itemize}}

\def\su5u1{SU(5) \times U(1)}
\def\fsu5u1{SU(5) \times U(1)'}
\def\so10{SO(10)}
\def\sq20{SO(10) \times SO(10)}

\begin{document}

\preprint{%
\vbox{%
\hbox{MADPH-04-1398}
\hbox{hep-ph/0410252}
\hbox{\today}
}}

\vspace*{.5in}

\title{Axion Models with High-Scale Supersymmetry Breaking}

\author{V. Barger}
\affiliation{Department of Physics, University of Wisconsin, 
Madison, WI 53706, USA}

\author{Cheng-Wei Chiang}
\affiliation{Department of Physics, University of Wisconsin, 
Madison, WI 53706, USA}
\affiliation{Department of Physics, National Central University, Chungli 320,
  Taiwan}

\author{Jing Jiang}
\affiliation{High Energy Physics Division, Argonne National
  Laboratory, Argonne, IL 60439, USA}
\affiliation{Institute of Theoretical Science, University of Oregon, 
Eugene, OR 97403, USA}

\author{Tianjun Li}
\affiliation{School of Natural Science, Institute for Advanced Study,
 Princeton, NJ 08540, USA
 \vspace*{.5in}}

\begin{abstract}
  
  Inspired by the possibility of high-scale supersymmetry breaking in
  the string landscape where the cosmological constant problem
  and the gauge hierarchy problem can be solved while the strong CP 
  problem is still a challenge for naturalness, we propose a 
  supersymmetric KSVZ axion model with an approximate universal 
  intermediate-scale ($\sim 10^{11}$~GeV) supersymmetry breaking. 
  To protect the global Peccei--Quinn (PQ) symmetry against quantum 
  gravitational violation, we consider the gauged discrete $Z_N$ PQ 
  symmetry.  In our model the axion can be a cold dark matter 
  candidate, and the intermediate supersymmetry breaking scale is 
  directly related to the PQ symmetry breaking scale. Gauge coupling 
  unification can be achieved at about $2.7\times 10^{16}$~GeV.
  The Higgs mass range is 130~GeV to 160~GeV.   We briefly discuss 
  other axion models with high-scale supersymmetry
  breaking where the stabilization of the axion solution is similar.

\end{abstract}

\pacs{}
\maketitle


\section{Introduction}

There exists a great variety of structures for the superstring/M theory vacua,
the ensemble of which is called the stringy ``landscape''~\cite{String}.  With
the ``weak anthropic principle''~\cite{Weinberg}, this proposal may not only
give the first concrete explanation of the very tiny value of the cosmological
constant which can only take discrete values, but also solve the gauge hierarchy
problem.  In particular, the supersymmetry breaking scale can be high if there
exist many supersymmetry breaking parameters or many hidden
sectors~\cite{HSUSY,NASD}.  Although there is no definite conclusion that
the string 
landscape predicts high-scale or TeV-scale supersymmetry breaking~\cite{HSUSY},
it is interesting to study models with high-scale supersymmetry
breaking~\cite{NASD}.

If supersymmetry is indeed broken at a high scale, the breaking scale can range
from 1 TeV to the string scale. For simplicity, we consider three
representative cases: (1)~string-scale supersymmetry breaking,
(2)~intermediate-scale supersymmetry breaking, and (3)~TeV-scale supersymmetry
breaking. The TeV-scale supersymmetry has been studied extensively during the
last two decades, so we do not discuss it here.

Although the original motivation for the New Minimal Standard Model
(NMSM)~\cite{Davoudiasl:2004be} is different from the above
 string landscape argument,
the NMSM provides a concrete model with string-scale supersymmetry breaking.
With only six extra degrees of freedom beyond the minimal Standard Model (SM),
the NMSM incorporates the new discoveries of physics beyond the minimal
Standard Model: dark energy, non-baryonic dark matter, neutrino masses, as well
as baryon asymmetry and cosmic inflation, by adopting the principle of minimal
particle content and the most general renormalizable Lagrangian.  The NMSM is
free from excessive flavor-changing effects, CP violation, too-rapid proton
decay, problems with electroweak precision data, and unwanted cosmological
relics~\cite{Davoudiasl:2004be}.  Moreover, gauge coupling unification can be
achieved by considering threshold corrections~\cite{Calmet:2004ck}, or by
introducing suitable additional particles, similar to the axion models with
string-scale supersymmetry breaking that will be discussed in Section~V.
The unitarity constraints on couplings are weaker than the stability and
triviality constraints~\cite{Cynolter:2004cq}.

For high-scale supersymmetry breaking, Arkani-Hamed and Dimopoulos proposed the
split supersymmetry scenario where the scalars (squarks, sleptons and one
combination of the scalar Higgs doublets) have masses at an
intermediate scale, while the fermions (gauginos and Higgsinos) and the other
combination of the scalar Higgs doublets are still at the TeV
scale~\cite{NASD}. Gauge coupling unification is preserved and the lightest
neutralino can still be a dark matter candidate. In addition, most of the
problems with supersymmetric models, for example,  flavor and CP
violations, dimension-5 fast proton decay and the stringent constraints on
the lightest Higgs mass, are solved. The consequences of split supersymmetry
have been studied in Refs.~\cite{GGAR,SPLIT}.

On the other hand, unlike the cosmological constant problem and the gauge
hierarchy problem, the strong CP problem is still a concern in the
string landscape~\cite{Donoghue:2003vs}.  In the Standard Model, the
$\overline{\theta} $ parameter is a dimensionless coupling constant which is
infinitely renormalized by radiative corrections.  There is no theoretical
reason for $\overline{\theta} $ to be smaller than $10^{-9}$ as required by the
experimental bound on the electric dipole moment of the neutron~\cite{review,
  pdg}.  There is also no known anthropic constraint on the value of
$\overline{\theta} $, {\it i.e.}, $\overline{\theta} $ may be a random variable
with a roughly uniform distribution in the string
landscape~\cite{Donoghue:2003vs}. An elegant and popular solution
to the strong CP problem is provided by the Peccei--Quinn (PQ)
mechanism~\cite{PQ}, in which a global axial symmetry $U(1)_{PQ}$ is
introduced and broken spontaneously at some high energy scale. Then
$\overline{\theta}$ is promoted to a dynamical field,
$\overline{\theta}=a/f_a$, with an effective potential for this field induced
by non-perturbative QCD effects.  The axion $a$ is a pseudo-Goldstone boson
from the spontaneous $U(1)_{PQ}$ symmetry breaking, with a decay constant $f_a$
(of similar scale as the $U(1)_{PQ}$ symmetry breaking scale).  Minimization of the
axion potential fixes the vacuum expectation value (VEV) of $a$, 
equivalently forces $\overline{\theta}$ to be zero, thus naturally solving the
strong CP problem.

The original Weinberg--Wilczek axion~\cite{WW} is excluded by experiment, in
particular by the non-observation of the rare decay $K \rightarrow \pi +
a$~\cite{review}.  There are two viable ``invisible'' axion models in which the
experimental bounds can be evaded: (1)~the Kim--Shifman--Vainshtein--Zakharov
(KSVZ) axion model, which introduces a SM singlet and a pair of extra
vector-like quarks that carry $U(1)_{PQ}$ charges while the SM fermions and
Higgs fields are neutral under $U(1)_{PQ}$ symmetry~\cite{KSVZ}; (2)~the
Dine--Fischler--Srednicki--Zhitnitskii (DFSZ) axion model, in which a
SM singlet and one pair of Higgs doublets are introduced,
and the SM fermions and Higgs fields
are charged under $U(1)_{PQ}$ symmetry~\cite{DFSZ}.

From laboratory, astrophysics, and cosmology constraints, the $U(1)_{PQ}$
symmetry breaking scale $f_a$ is limited to  the range $10^{10}~{\rm GeV}
\leq f_a \leq 10^{12}~{\rm GeV}$~\cite{review}.  Light axions can be
produced in stars, and part of the star energy can be carried away by these
axions.  To avoid an unacceptable star energy loss, a lower bound is
obtained on the axion decay constant, $f_a \geq 10^{10}$~GeV.
During the course of cosmological evolution of the universe, the axions
decouple early and begin to oscillate coherently. If $f_a$ is larger than about
$10^{12}$~GeV, at some point in the evolution the energy density in the
coherent axion oscillations could exceed the critical energy density and
over-close the universe.  The invisible axion can be a good cold dark matter
candidate~\cite{review}.

In this paper we consider the axion models with high-scale supersymmetry breaking.
Unlike split supersymmetry, we consider an approximately universal supersymmetry
breaking, {\it i.e.}, the mass differences among the squarks, sleptons,
gauginos, Higgsinos and the $A$-terms are within one order of magnitude (within
one-loop suppressions), which seems to be quite natural from flux-induced
supersymmetry breaking~\cite{Camara:2004jj}.

We construct a supersymmetric KSVZ axion model where we introduce a SM singlet
$S$ and two pairs of SM vector-like particles ($Q_X$, $\overline{Q}_X$) and
($D_X$, $\overline{D}_X$).  The supersymmetry breaking scale and the
Peccei--Quinn symmetry breaking scale are both around $10^{11}$~GeV, and the
axion can be a cold dark matter candidate.  We give the Higgs potential and
the SM Yukawa couplings below the intermediate scale, the superpotential above
the intermediate scale, and the matching conditions for the couplings.  In
addition, we consider the discrete $Z_N$ Peccei--Quinn symmetry, which can be
embedded into an anomalous $U(1)_A$ gauge symmetry in string constructions
where the anomalies can be cancelled by the Green--Schwarz
mechanism~\cite{MGJS}.  This $Z_N$ discrete symmetry cannot be violated by the
quantum gravitational interaction.  In order that the contributions to the
$\overline{\theta}$ term from the non-renormalizable operators can be less than
$10^{-9}$, we must require $N \geq ~ 10$.  We also show that a suitable
quartic coupling for the singlet $S$ can be obtained at one-loop due to its
large Yukawa couplings to the extra vector-like particles.  Thus, the
intermediate supersymmetry breaking scale is directly connected to the
Peccei--Quinn symmetry breaking scale. Moreover, using two-loop renormalization
group equation runnings for the gauge couplings and one-loop renormalization
group equation runnings for the Yukawa couplings, we find that gauge coupling
unification can be achieved at about $2.7\times 10^{16}$~GeV and the Higgs
mass range is from 130~GeV to 160~GeV.  Also, we calculate the Yukawa couplings
for the third family of the SM fermions at the Grand Unified Theory (GUT)
scale and comment on how to achieve Yukawa coupling unification.  In our
model, similar to the NMSM and split supersymmetry, most of the usual problems
in the supersymmetric models are solved.  If the $R$-parity is an exact symmetry,
the lightest supersymmetric particle (LSP) neutralinos can be produced only
non-thermally in a suitable amount, and their annihilations may have
cosmological consequences.  If $R$-parity is broken, the tiny masses and
bilarge mixings for active neutrinos can be realized naturally at one-loop
although we have to suppress the contributions to neutrino masses from the
tree-level $\mu_i L_i H_u$ terms in the superpotential.

We also briefly discuss other axion models: the DFSZ axion model with
intermediate-scale supersymmetry breaking where the stabilization of the axion
solution and the Peccei--Quinn symmetry breaking
are similar, the axion models with string-scale supersymmetry breaking, and
the axion models with split supersymmetry.

\section{The KSVZ Axion Model with Intermediate-Scale\\ Supersymmetry Breaking} 
\label{sec:model}

We first specify our conventions. We denote the left-handed quark doublets, the
right-handed up-type quarks, the right-handed down-type quarks, the left-handed
lepton doublets, the right-handed leptons, and the Higgs doublet in the Standard
Model as $q_i$, $u_i$, $d_i$, $l_i$, $e_i$ and $H$, respectively, where $i$ is
the family index.  For the supersymmetric Standard Model, the SM fermions and
Higgs fields are superfields belonging to chiral multiplets.  We denote the
left-handed quark doublets, the right-handed up-type quarks, the right-handed
down-type quarks, the left-handed lepton doublets, the right-handed leptons, and
one pair of Higgs doublets as $Q_i$, $U^c_i$, $D^c_i$, $L_i$, $E^c_i$, $H_u$
and $H_d$, respectively.  The supersymmetry
breaking scale is assumed to be at an intermediate scale in the range
$10^{10}$~GeV to $10^{12}$~GeV. For simplicity, we assume that the gauginos,
squarks, sleptons, Higgsinos, and one combination of the scalar Higgs doublets
have a universal supersymmetry breaking soft mass ${\widetilde M}$.

We consider the KSVZ axion model~\cite{KSVZ} with a discrete Peccei--Quinn
symmetry $Z_N$. This $Z_N$ symmetry can be realized after the anomalous
$U(1)_A$ gauge symmetry is broken close to the string scale, where the $U(1)_A$
anomalies are cancelled by the Green--Schwarz mechanism~\cite{MGJS}, as will
be discussed in the next Section.  In the KSVZ axion model, the SM fermions and
Higgs fields are assumed to be neutral under the $Z_N$ PQ symmetry.  We
introduce one SM singlet $S$ and two pairs of SM vector-like particles
($Q_X$, $\overline{Q}_X$) and ($D_X$, $\overline{D}_X$).  Their quantum numbers
under the SM $SU(3)_C\times SU(2)_L\times U(1)_Y$ gauge symmetry and the $Z_N$
symmetry are specified in Table~\ref{QNEP}.  The axion potential from the QCD
anomaly is generated when we couple the singlet $S$ to these extra vector-like
particles.  In addition, the charges of $Q_X$, $\overline{Q}_X$, $D_X$,
$\overline{D}_X$, and $S$ under the $Z_N$ symmetry satisfy the relations
\begin{eqnarray}
n_{Q_X} + n_{\overline{Q}_X} + n_S & = & 0 ~{\rm mod}~N~,~\,
\end{eqnarray}
\begin{eqnarray}
n_{D_X} + n_{\overline{D}_X} + n_S  &=& 0 ~{\rm mod}~N~.~\,
\end{eqnarray}

\begin{table}[t]
\caption{ The quantum numbers for the superfields $Q_X$, $\overline{Q}_X$,
$D_X$, $\overline{D}_X$, and $S$  under the
$SU(3)_C\times SU(2)_L\times U(1)_Y\times Z_N$ symmetry.
 \label{QNEP}} 
\vspace{0.4cm}
\begin{center}
\begin{tabular}{|c|c|c|c|}
\hline
 Superfields &  Quantum Numbers & Superfields &  Quantum Numbers \\
\hline
$Q_X$ & ({\bf 3}, {\bf 2}, {\bf 1/6}, ${\bf n_{Q_X}}$) & 
$\overline{Q}_X$ & (${\bf \overline{3}}$, {\bf 2}, {\bf $-$1/6}, ${\bf
 n_{\overline{Q}_X}}$) \\
\hline 
$D_X$ & ({\bf 3}, {\bf 1}, {\bf $-$1/3}, ${\bf n_{D_X}}$) & 
$\overline{D}_X$ & (${\bf \overline{3}}$, {\bf 1}, {\bf 1/3}, $ {\bf
 n_{\overline{D}_X}}$) \\
\hline 
$S$ & ({\bf 1}, {\bf 1}, {\bf 0}, ${\bf n_{S}}$) & & \\
\hline
\end{tabular}
\end{center}
\end{table}

In our model the two pairs of the SM vector-like particles ($Q_X$,
$\overline{Q}_X$) and ($D_X$, $\overline{D}_X$) have masses comparable to the
universal supersymmetry breaking soft mass ${\widetilde M}$ after the
supersymmetry and Peccei-Quinn symmetry breakings.  Below the scale
${\widetilde M}$ we assume that only one scalar Higgs doublet $H$ is light as
arranged by suitable fine-tuning.  Thus, besides the gauge kinetic terms for
all the fields, the most general renormalizable Lagrangian at tree-level is
\begin{eqnarray}
{\cal L}&=& m_H^2 H^\dagger H-\frac{\lambda}{2!}\left( H^\dagger H\right)^2
-\left[ h^u_{ij} {\bar q}_j u_i\epsilon H^* 
+h^d_{ij} {\bar q}_j d_iH
+h^e_{ij} {\bar \ell}_j e_iH + {\rm H.C.}\right]~,~\,
\label{Lagr}
\end{eqnarray}
where $m_H^2$ is the squared Higgs mass, $\lambda$ is the Higgs quartic
coupling, $\epsilon \equiv i \sigma^2 $ where $\sigma^2$ is the second Pauli
matrix, and $h^u_{ij}$, $h^d_{ij}$, and $h^e_{ij}$ are the Yukawa couplings.

The Lagrangian in Eq.~(\ref{Lagr}) is obtained after the gauginos, squarks,
sleptons, Higgsinos, and one combination of the scalar Higgs doublets from
$H_u$ and $H_d$ in the supersymmetric model are decoupled at the scale
$\widetilde{M}$.  Matching the Lagrangian in Eq.~(\ref{Lagr}) with the 
supersymmetric interaction terms for the Higgs doublets $H_u$ and $H_d$, we
obtain the relevant Lagrangian in terms of the coupling constants in the SM
at the scale $\widetilde{M}$
\begin{eqnarray}
{\cal L}_{\rm SUSY}&=&
-\frac{g_2^2}{8}\left( H_u^\dagger \sigma^i H_u + H_d^\dagger \sigma^i H_d
\right)^2
-\frac{g_Y^{ 2}}{8}\left( H_u^\dagger H_u - H_d^\dagger  H_d
\right)^2 \nonumber \\
&&+\left[ y^u_{ij}H_u \epsilon {\bar u}_i q_j
-y^d_{ij}H_d \epsilon {\bar d}_i q_j
-y^e_{ij}H_d \epsilon {\bar e}_i \ell_j + {\rm H.C.} \right]~,~\,
\label{LagrSUSY}
\end{eqnarray}
where $\sigma^i$ are the Pauli matrices, $g_2$ and $g_Y$ are respectively the
gauge couplings for the $SU(2)_L$ and $U(1)_Y$ groups, and $y^u_{ij}$,
$y^d_{ij}$, and $y^e_{ij}$ are the Yukawa couplings at and above scale
$\widetilde M$.  Here we neglect the Yukawa couplings for the extra vector-like
particles that will be given later in Eq.~(\ref{SuperP}).  

From the scalar Higgs
doublets in the chiral multiplets, $H_u$ and $H_d$, we define the SM Higgs
doublet combination as $H \equiv -\cos\beta \epsilon H_d^*+\sin\beta H_u$ which
is fine-tuned to have a small mass term, while the other combination
($\sin\beta \epsilon H_d^*+\cos\beta H_u $) has mass around ${\widetilde
  M}$~\cite{NASD}. Therefore, we obtain the coupling constants in
Eq.~(\ref{Lagr}) at the scale $\widetilde{M}$ from those in
Eq.~(\ref{LagrSUSY}) by the replacements $H_u\to \sin\beta H$ and $H_d\to
\cos\beta \epsilon H^*$
\begin{eqnarray}
\lambda(\widetilde M )& =& \frac{\left[ g_2^2(\widetilde M )+3g_1^{
      2}(\widetilde M)/5 \right]}{4} \cos^22\beta ~,~\,
\label{MatchH}
\end{eqnarray}
\begin{eqnarray}
h^u_{ij}(\widetilde M )=(y^{u}_{ij})^*(\widetilde M )\sin\beta , &&
h^{d,e}_{ij}(\widetilde M )=(y^{d,e}_{ij})^*(\widetilde M )\cos\beta ~,~\,
\label{MatchY}
\end{eqnarray}
where we assume the $SU(5)$ normalization for the gauge coupling of $U(1)_Y$,
{\it i.e.}, $g_1^2 \equiv (5/3)g_Y^{ 2}$.

Above the scale ${\widetilde M}$ the theory is supersymmetric.  The
superpotential in our model is
\begin{eqnarray}
W &=& y^u_{ij} H_u \epsilon U^c_i Q_j
-y^d_{ij} H_d \epsilon D^c_i Q_j
-y^e_{ij} H_d \epsilon E^c_i L_j
+y_{Q_X} S Q_X \overline{Q}_X 
\nonumber \\ &&
+ y_{D_X} S D_X \overline{D}_X +
 y_S {{S^N}\over\displaystyle {M_{Pl}^{N-3}}} ~,~\,
\label{SuperP}
\end{eqnarray}
where $y_{Q_X}$, $y_{D_X}$ and $y_S$ are the Yukawa couplings for the extra
particles, and $M_{Pl}$ is the Planck scale.  The last term, $y_S
S^N/M_{Pl}^{N-3}$, is the lowest possible higher dimensional operator for the
singlet $S$; this operator is related to the destabilization of the axion
solution and suppressed by the Planck scale. Consequently, in the following
discussions we do not consider the renormalization group evolution for the
coupling $y_S$ and neglect it in our renormalization group equations.

\section{Stabilization of the Axion Solution and Peccei--Quinn\\ Symmetry
  Breaking}
\label{sec:stab}

Quantum gravitational effects, associated with black holes, worm holes, etc.,
are believed to violate all the global symmetries, while they respect all the
gauge symmetries~\cite{Hawking}.  These effects may destabilize the axion
solution to the strong CP problem due to the violation of the global Peccei--Quinn
symmetry.  However, after a gauge symmetry is spontaneously broken, there may
exist a remnant discrete gauge symmetry which will not be violated by quantum
gravity~\cite{LKFW}.  Thus, a possible way to avoid the destabilization problem
associated with quantum gravity is to identify the Peccei--Quinn symmetry as an
approximate global symmetry arising from the broken gauge symmetry.

In weakly coupled heterotic string model building, there generically exists an
anomalous $U(1)_A$ gauge symmetry with its anomalies cancelled by the
Green--Schwarz mechanism~\cite{MGJS}.  For Type II orientifold string model
building, there may exist more than one anomalous $U(1)_A$ gauge symmetry
whose anomalies can be cancelled by the generalized Green--Schwarz
mechanism~\cite{Aldazabal:2000dg}.  The anomalous $U(1)_A$ gauge symmetry is
broken near the string scale when some scalar fields, which are charged under
$U(1)_A$, obtain VEVs and cancel the
Fayet--Iliopoulos term of $U(1)_A$. Then the D-flatness for $U(1)_A$ is
preserved and the supersymmetry is unbroken~\cite{DSW}.  Usually, there is an
unbroken discrete $Z_N$ subgroup of the $U(1)_A$ gauge symmetry, which is
protected against quantum gravitational violation.  We shall consider this
$Z_N$ discrete symmetry as an approximate global $U(1)_{PQ}$
symmetry~\cite{Babu:2002ic}.

For the gauge symmetry $\prod_i G_i \times U(1)_A$, the Green--Schwarz anomaly
cancellation conditions from an effective theory point of view are~\cite{TBMD,
  Ibanez}
\begin{eqnarray}
\frac{A_{i}}{k_{i}}=\frac{A_{gravity}}{12}=\delta_{GS}~,~\,
\end{eqnarray}
where the $A_i$ are anomaly coefficients associated with $G_i^2 \times U(1)_A$,
$k_i$ is the level of the corresponding Kac--Moody algebra, and $\delta_{GS}$
is a constant which is not specified by low-energy theory alone.  For a
non-Abelian group, $k_i$ is a positive integer, while for the $U(1)$ gauge
symmetry, $k_i$ need not be an integer.  All the other anomaly coefficients
such as $G_i G_j G_k$ and $[U(1)_A]^2 \times G_i$ should vanish.

In our model the gauge symmetry in which we are interested is $SU(3)_C \times
SU(2)_L \times U(1)_Y \times U(1)_A$. Because the SM fermions and Higgs fields
are not charged under $U(1)_A$, there are no anomalies from them.  Thus, we
only need to consider the anomalies from the two pairs of SM vector-like fields
($Q_X$, $\overline{Q}_X$) and ($D_X$, $\overline{D}_X$) that involve at least
one $U(1)_A$.

For simplicity, we assume that
$n_{Q_X} = n_{D_X}$ and $n_{\overline{Q}_X} =  n_{\overline{D}_X}$.
 Using the Georgi--Glashow $SU(5)$
normalization~\cite{Georgi:1974sy}, we find that the $U(1)_Y [U(1)_A]^2$
anomaly vanishes and that the $[U(1)_Y]^2 \times U(1)_A$, $[SU(2)_L]^2 \times
U(1)_A$ and $[SU(3)_C]^2 \times U(1)_A$ anomaly coefficients $A_1$, $A_2$ and
$A_3$ are
\begin{eqnarray}
5 A_1=A_2 = A_3={3\over 2} \left( n_{Q_X} + 
n_{\overline{Q}_X} \right)~.
\end{eqnarray}
Thus, the anomalies can be cancelled by choosing the $k_i$ as follows
\begin{eqnarray}
k_1={1\over 5}~,~ k_2=k_3=1~.~\,
\end{eqnarray}
Usually, one only considers non-Abelian anomaly cancellations because the
associated Kac--Moody level for $U(1)$ is not an integer in general, hence this
condition is not very useful from a low-energy effective theory point of
view~\cite{Babu:2002ic, TBMD}.

In addition, we can have models with $k_1=k_2=k_3=1$,
although we have to introduce more particles. 
Let us give an explicit example. To
ensure gauge coupling unification, we introduce two pairs of SM vector-like
fields with quantum numbers ((${\bf 3}, {\bf 2}, {\bf 1/6}$), (${\bf
  \overline{3}}$, ${\bf 2}$, ${\bf -1/6}$)) and ((${\bf 3}$, ${\bf 1}$, ${\bf
  -1/3}$), (${\bf \overline{3}}$, ${\bf 1}$, ${\bf 1/3}$)).  An
alternative way to achieve gauge coupling unification is to introduce two
chiral multiplets with quantum numbers (${\bf 8}, {\bf 1}, {\bf 0}$) and (${\bf
  1}, {\bf 3}, {\bf 0}$), and one pair of SM vector-like particles with
quantum numbers ((${\bf 1}$, ${\bf 2}$, ${\bf -1/2}$), (${\bf 1}$, ${\bf 2}$,
${\bf 1/2}$)). In these two cases,
all the extra particles are neutral under $U(1)_A$ and have masses around
$10^{11}$~GeV, and the gauge coupling unification scale is
about $2\times 10^{16}$~GeV.  To realize the KSVZ axion and preserve the gauge
coupling unification, we introduce at least one pair of SM vector-like fields
with $SU(5)$ quantum numbers, for instance, ($\bf 5$, $\bf \overline{5}$) or
($\bf 10$, $\bf \overline{10}$), that are charged under $U(1)_A$ and can
couple to the singlet $S$ via superpotential $S {\bf 5} {\bf \overline{5}}$ or
$S {\bf 10} {\bf \overline{10}}$ terms. Because these SM vector-like fields, which
couple to $S$ and give the axion potential from QCD anomaly, form a complete
$SU(5)$ representation, we have $k_1=k_2=k_3=1$. However, this model is not 
economical compared to above model because we need to add more
particles.

To ensure that the contributions to the $\overline{\theta}$ term from the
non-renormalizable operators are less than $10^{-9}$, we constrain the order
$N$ of the $Z_N$ symmetry. The most dangerous term is the following hard
supersymmetry breaking term in the potential:
\begin{eqnarray}
V_{SB} &=& A_{y_S} y_S 
{{S^N}\over\displaystyle {M_{Pl}^{N-3}}} +{\rm H.C.}~,~\,
\end{eqnarray}
where $A_{y_S}$ can be complex.  This term would induce a non-zero contribution
to $\overline{\theta}$~\cite{Barr} of
\begin{eqnarray}
\Delta \overline{\theta}~ \simeq ~ A_{y_S} y_S 
\frac{f_a^{N}}{{M_{ Pl}}^{N-3}\Lambda_{QCD}^{4}}~.
\end{eqnarray}
Here we assumed that CP violation is of order 1 so that there is no extra
fine-tuning in our discussion.

For numerical calculations, we first evaluate $f_a$.  Assuming that the ratio
of the axion number density to the entropy density has been constant since the
axion acquired mass and started to oscillate, we have~\cite{Sikivie:1999sy}
\begin{eqnarray}
\Omega_a h^2 \simeq \left( {{f_a}\over\displaystyle
{10^{12} ~{\rm GeV}}} \right)^{7/6}
\left( {{200 ~{\rm MeV}}\over\displaystyle
{\Lambda_{QCD}}} \right)^{3/4} ~.~\,
\end{eqnarray}
Using the dark matter relic density of $\Omega_a h^2 \simeq 0.1126$
from the WMAP analysis~\cite{Spergel:2003cb} and taking $\Lambda_{QCD}
=200$~MeV, we obtain $f_a \simeq 1.54 \times 10^{11}$~GeV.

Assuming that $y_S=1$ and requiring that $\Delta \overline{\theta} < 10^{-9}$,
we obtain
\begin{eqnarray}
 \left\{ \begin{array}{ll}
N \geq 11, & ~~~~~~ {\rm for}~~ A_{y_S} \sim 10^{11} ~{\rm GeV} \\
N \geq 10, & ~~~~~~ {\rm for}~~ A_{y_S} \sim 10^{4} ~{\rm GeV} 
 \end{array} \right. \ ~,~\,
\end{eqnarray}
where we take $M_{Pl} =2.4\times 10^{18}$~GeV.  If the CP violation phase 
or the magnitude of $A_{y_S}$,
or the coupling $y_S$ are very small, the values $N$ can be smaller than
these lower bounds.

The above stabilization discussion is generic for supersymmetric axion models
because the hard supersymmetry breaking terms can be induced from the
non-renormalizable operators in the K\"ahler potential after supersymmetry
breaking.  For example, in our model we can have the following
non-renormalizable operator in the K\"ahler potential
\begin{eqnarray}
S &=& \int d^4x d^4\theta \left(
{{y_{ZS} Z^{\dagger} Z S^N} \over\displaystyle {M_*^N}}
+ {\rm H.C.} \right) ~,~\,
\end{eqnarray}
where $y_{ZS}$ is a coupling constant, and
$Z$ is a spurion superfield which parametrizes the supersymmetry breaking
via $\langle F_Z \rangle \not=0$.  For simplicity, we assume that the mass
scale $M_*$ is equal to the Planck scale $M_{Pl}$ and $\langle F_Z
\rangle/M_{Pl}$ is about $10^{11}$ GeV.  Therefore, we obtain
\begin{eqnarray}
A_{y_S} = {{y_{ZS}}\over {y_S}} {{|F_Z|^2}\over {M_{Pl}^3}}
\sim  {{y_{ZS}}\over {y_S}} \times 10^4 ~{\rm GeV}~.~\,
\end{eqnarray}
Using $y_{ZS} \sim 1$ and $y_S \sim 1$, we obtain $N \geq 10$.

In general, suppose we have several singlets $S_i$ that are charged under the
$Z_N$ symmetry whose VEVs are comparable to that of $S$, and suppose that the
lowest dimensional operator allowed by the gauge symmetry and $Z_N$ symmetry is
$S^{n_0} \prod_i S_i^{n_i} $, where $n_0$ is a positive integer and $n_i$ are
non-negative integers.  To ensure that the contribution to the
$\overline{\theta}$ term from this operator is naturally less than $10^{-9}$,
we require
\begin{eqnarray}
n_0+\sum_i n_i \geq 11 ~{\rm or}~ 10~.~\,
\end{eqnarray}

We next discuss Peccei--Quinn symmetry breaking.  Because the soft mass and the
VEV of the singlet $S$ are about $10^{11}$~GeV, we have to generate a suitable
quartic coupling for $S$ at one-loop from the two pairs of SM vector-like
fields. This can be realized as long as $y_{Q_X}$ and $y_{D_X}$ are of order 1.
The one-loop effective potential for $S$ is
\begin{eqnarray}
V_{eff} &=& -{\tilde m}^2_S |S|^2 
-{3\over {8\pi^2}} y_{Q_X}^4 |S|^4
\left( {\rm log} {{ y_{Q_X}^2 |S|^2}
\over {Q^2}} -{3\over 2} \right)
-{3\over {16\pi^2}} y_{D_X}^4 |S|^4
\left( {\rm log} {{ y_{D_X}^2 |S|^2}
\over {Q^2}} -{3\over 2} \right)
\nonumber \\ &&
{}+{3\over {8\pi^2}} \left({\tilde m}_{Q_X}^2 
+ y_{Q_X}^2 |S|^2 \right)^2
\left( {\rm log} {{{\tilde m}_{Q_X}^2 + y_{Q_X}^2 |S|^2 }
\over {Q^2}} -{3\over 2} \right)
\nonumber \\ &&
{}+{3\over {16\pi^2}} \left({\tilde m}_{D_X}^2 
+ y_{D_X}^2 |S|^2 \right)^2
\left( {\rm log} {{{\tilde m}_{D_X}^2 + y_{D_X}^2 |S|^2 }
\over {Q^2}} -{3\over 2} \right) ~,~\,
\end{eqnarray}
where ${\tilde m}^2_S$ is the soft supersymmetry breaking mass for $S$,
${\tilde m}_{Q_X}^2$ is the soft mass for $ Q_X$ and $\overline{Q}_X$, and
${\tilde m}_{D_X}^2$ is the soft mass for $ D_X$ and $\overline{D}_X$;  $Q$ is
the renormalization scale, which is about $f_a$ here.  Approximately, the
quartic term for $S$ is
\begin{eqnarray}
V_{eff} \supset {{\lambda_S}\over {2!}}
(S^{\dagger}S)^2 ~,~\,
\end{eqnarray}
where 
\begin{eqnarray}
\lambda_S =
{3\over {4\pi^2}} y_{Q_X}^4 
 {\rm log} \left( {{{\tilde m}_{Q_X}^2 }
\over {y_{Q_X}^2 f_a^2}} +1 \right)
+ {3\over {8\pi^2}} y_{D_X}^4 
 {\rm log} \left( {{{\tilde m}_{D_X}^2 }
\over {y_{D_X}^2 f_a^2}} +1 \right) ~.~\,
\end{eqnarray}
For $y_{Q_X}=y_{D_X}=0.8$ and ${\tilde m}_{Q_X} = {\tilde m}_{D_X} = 5 f_a$,
we obtain $\lambda_S =0.17$.

We emphasize that for a supersymmetry breaking scale higher 
than $10^{12}$ GeV, the VEV of $S$ around
$f_a \sim 1.54\times 10^{11}$ GeV may not be natural. For a supersymmetry
breaking scale lower than $10^{10}$ GeV, one may not be able to generate the
suitable quartic coupling for the singlet $S$ naturally. Therefore,
the Peccei--Quinn symmetry can be broken naturally 
if and only if the supersymmetry breaking scale
is intermediate, {\it i.e.}, from $10^{10}$ GeV to $10^{12}$ GeV, and then
the intermediate supersymmetry breaking scale is directly
related to the Peccei--Quinn symmetry breaking scale.

\section{ Gauge Coupling Unification and Phenomenology}

\subsection{Gauge Coupling Unification}

We next consider gauge coupling unification.  The two-loop renormalization
group equations for the gauge couplings and the one-loop renormalization group
equations for the Yukawa couplings are given in Appendix A.  The
renormalization group equation runnings are logarithmic and we shall neglect the
uncertainties from threshold corrections. For simplicity, we consider the
universal supersymmetry breaking mass ${\widetilde M}$ for the superpartners of
the SM particles, and we assume that the masses for the two pairs of SM
vector-like superfields ($Q_X$, $\overline{Q}_X$) and ($D_X$, $\overline{D}_X$)
are the same and given by $m_{X}$.

In numerical calculations we use the following initial values at the $M_Z$
scale in Ref.~\cite{Eidelman:2004wy}:
\bea
&&\alpha^{-1}(M_Z) = 128.91 \pm 0.02\,, \nonumber \\
&&\sin^2\theta_W(M_Z) = 0.23120 \pm 0.00015\,, \nonumber \\
&&M_Z = 91.1876 \pm 0.0021\,{\rm GeV}\,, \nonumber \\
&&v = 174.10\,{\rm GeV}\,,
\eea
along with a top quark pole mass $m_t = 178.\pm 4.3$ GeV~\cite{Azzi:2004rc} and
$\alpha_s(M_Z) = 0.1182 \pm 0.0027$~\cite{Bethke:2004uy}.  We choose
$m_{X} = 1.0\times 10^{11}$ GeV, and $y_{Q_X} = y_{D_X} =0.8$ at the $m_X$
scale.  The supersymmetry breaking scale $\widetilde M$ and the GUT scale
$M_{GUT}$ are determined by requiring the gauge couplings to unify at
$M_{GUT}$. For the ${\overline{MS}}$ top quark Yukawa coupling, we use the one-loop
corrected value~\cite{Arason:1991ic}, which is related to the top quark pole
mass by
\be
m_t = h_t v \left(1 + \frac{16}{3}\frac{g_3^2}{16\pi^2} - 2\frac{h_t^2}{16\pi^2}\right)\,.
\ee

\begin{figure}[htb]
\centering
\includegraphics[width=8.0cm]{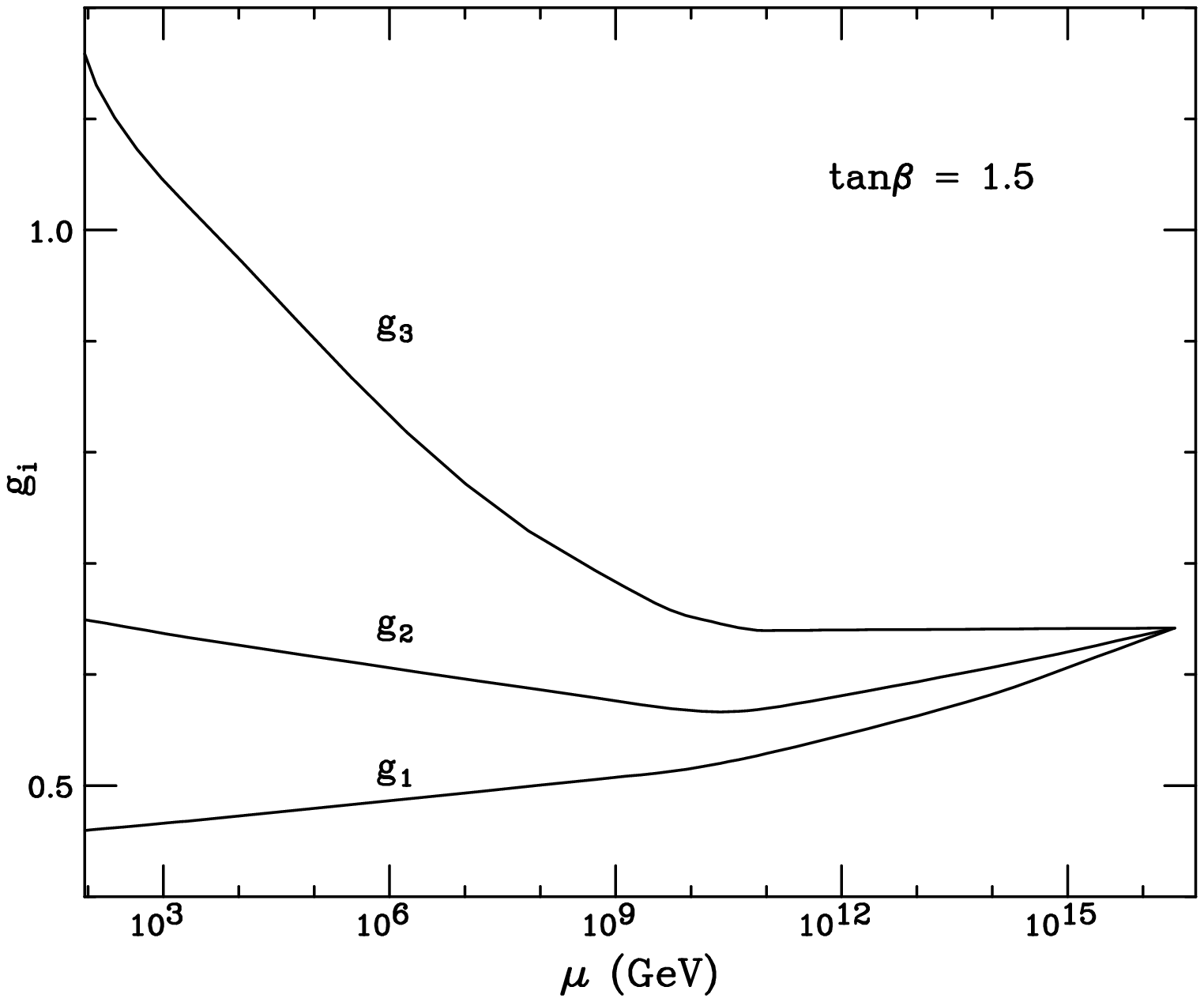}
\includegraphics[width=8.0cm]{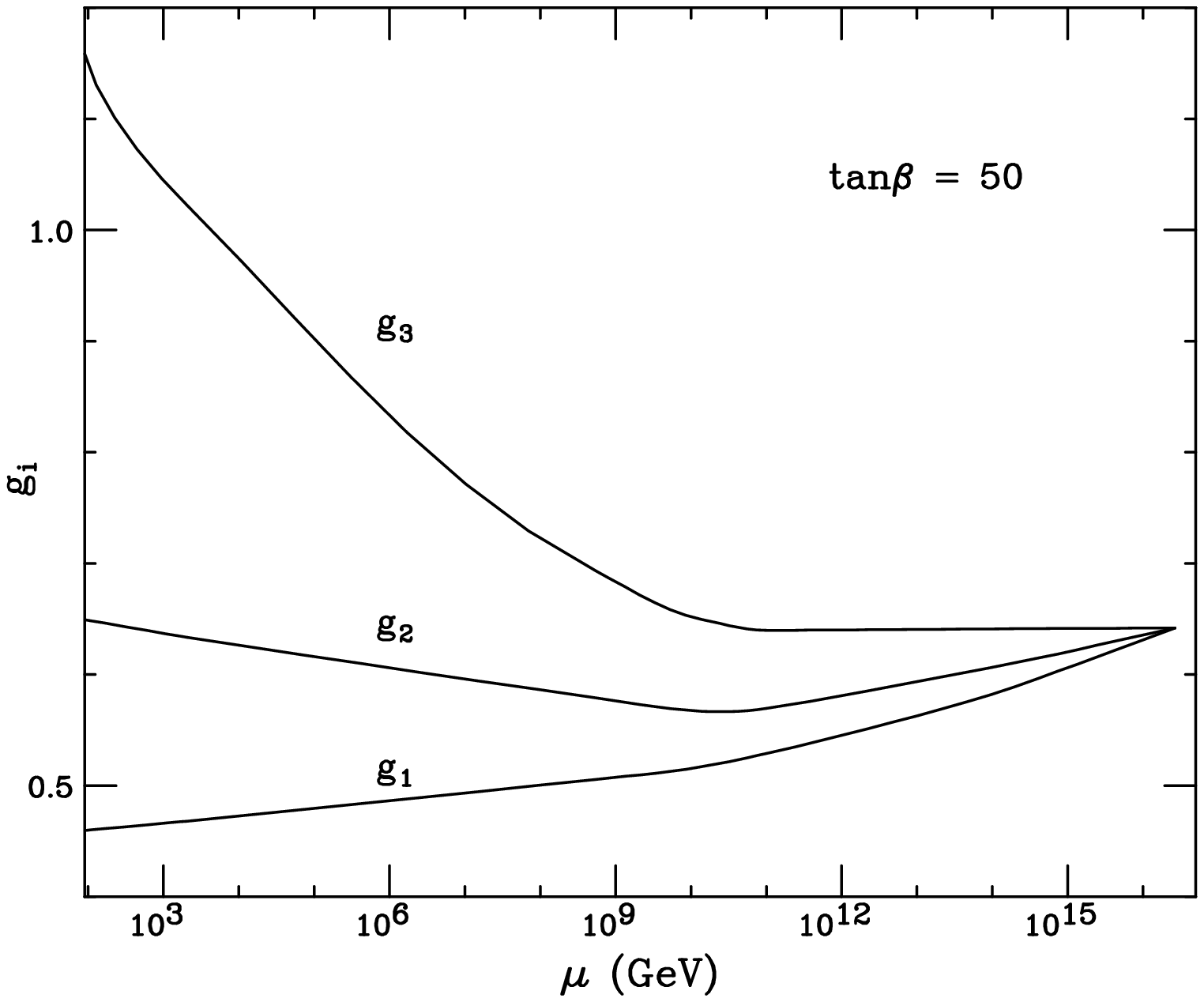}
\caption{Gauge coupling unifications  for $\tan\beta = 1.5$ (left) and 
$\tan\beta = 50$ (right).}
\label{fig:unilo}
\end{figure}

The unifications of the gauge couplings for $\tan\beta = 1.5$ and $\tan\beta =
50$ are shown in Fig.~\ref{fig:unilo}, in which the supersymmetry breaking
scale $\widetilde M$ is $1.4 \times 10^{10}$ GeV. The GUT scale $M_{GUT}$ is
about $2.6 \times 10^{16}$ GeV and $2.7 \times 10^{16}$ GeV for $\tan\beta =
1.5$ and $\tan\beta = 50$, respectively.  The supersymmetry breaking scale, the
GUT scale, and the unified gauge coupling are almost independent of
$\tan\beta$.

We point out that if the gauginos and Higgsinos have masses of about $\widetilde M$ while
the squarks and sleptons are about one order of magnitude heavier, the gauge
coupling unification scale is the same as above because the squarks and
sleptons form complete $SU(5)$ representations.

\subsection{Higgs Mass}

\begin{figure}[htb]
\centering
\includegraphics[width=8.0cm]{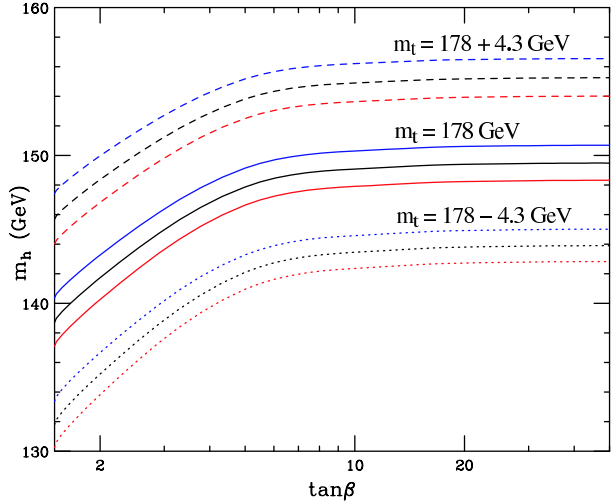}
\caption{The Higgs mass versus $\tan\beta$. The group of solid curves are 
	results for $m_t = 178.0$~GeV, the dashed for $m_t = 178.0+4.3$~GeV, and 
	the dotted for $m_t = 178.0-4.3$~GeV.  Within each group, the center curve 
	is generated with $\alpha_s = 0.1182$, the lower curve with 
	$\alpha_s = 0.1182 + 0.0027$, and the upper with 
	$\alpha_s = 0.1182 - 0.0027$.}
\label{fig:hmass}
\end{figure}

To calculate the Higgs mass, we first evaluate the Higgs quartic coupling
$\lambda$ at the scale $\widetilde M$ and then evolve it down to the $M_Z$
scale.  The effective potential, with the top quark radiative correction
included, is
\be
V_{eff} = m_h^2 H^\dagger H - \frac{\lambda}{2!} (H^\dagger H)^2 -
\frac{3}{16\pi^2} h_t^4 (H^\dagger H)^2 \left[\log\frac{h_t^2 (H^\dagger
H)}{Q^2} - \frac{3}{2}\right]\,.
\ee
We derive the Higgs mass by minimizing the effective potential. The scale $Q$
is chosen to be at the Higgs mass.  We show the predicted Higgs mass as a
function of $\tan\beta$ in Fig.~\ref{fig:hmass}.  The upper, center and lower
groups of curves correspond to the top quark mass being $(178.0+4.3)$~GeV,
$178.0$~GeV and $(178.0-4.3)$~GeV, respectively.  Within each group, the upper and
lower curves show the uncertainty due to the $1\sigma$ uncertainty of
$\alpha_s$.  The lower curve is generated with $\alpha_s = 0.1182 + 0.0027$ and
the upper with $\alpha_s = 0.1182 - 0.0027$.

Below the scale $\widetilde M$, the Higgs boson is just as in the SM.
A SM Higgs boson in the mass range 130--160~GeV can be identified at
the $3\sigma$ level of significance with $30~{\rm fb}^{-1}$ integrated
luminosity at the Tevatron~\cite{Han:1998ma,Carena:2000yx}.  At the
LHC, the Higgs boson in this mass range can be discovered in the
$\gamma\gamma$ and the $\tau^+\tau^-$ channels with $100~{\rm
fb}^{-1}$ of data~\cite{Higgs_LHC}.

\subsection{Top, Bottom and Tau Yukawa Couplings at the GUT Scale}

\begin{figure}[htb]
\centering
\includegraphics[width=8.0cm]{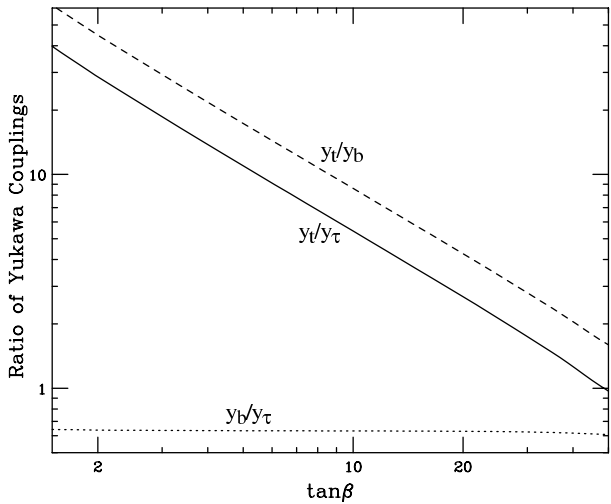}
\caption{The ratios of Yukawa couplings at the GUT scale
versus $\tan\beta$.  
The solid curve is $y_t/y_{\tau}$, 
the dashed $y_t/y_b$, and the dotted $y_b/y_{\tau}$.} 
\label{fig:yrat}
\end{figure}

Besides gauge coupling unification, it is also interesting to study whether
there exists Yukawa coupling unification for the top ($t$) quark, the
bottom ($b$) quark and the $\tau$ lepton at the GUT scale.  The ratios of the
Yukawa couplings ($y_t/y_{\tau}$, $y_t/y_b$ and $y_b/y_{\tau}$) as a function of
$\tan\beta$ are shown in Fig.~\ref{fig:yrat}.

From Fig.~\ref{fig:yrat}, we find that the $t$--$\tau$ Yukawa couplings would be
unified at $\tan\beta \simeq 49$, while the $t$--$b$ or $b$--$\tau$ Yukawa
couplings cannot be unified. Thus, how to explain the $t$, $b$, and $\tau$
Yukawa couplings at the GUT scale remains an interesting problem.  This may be
solved via flavor symmetry breaking or via particle mixings by introducing
additional SM vector-like particles with similar SM quantum numbers.

\subsection{Comments}

In our models, the gauge coupling unification scale is about $2.7\times
10^{16}$~GeV; therefore, there is no dimension-6 proton decay problem. Due to
intermediate-scale supersymmetry breaking, we do not have the dimension-5
proton decay problem, the moduli problem, and the excessive supersymmetry
flavor violation and CP violation, etc.  In addition, if we want to explain
neutrino masses via the see-saw mechanism~\cite{seesaw} and baryon asymmetry
via leptogenesis~\cite{LG86}, we just need to introduce three right-handed
neutrinos.  Also, the SM fermion masses and mixings can be generated naturally
via the Froggatt--Nielsen mechanism~\cite{FN}.

If $R$-parity is conserved in our model, the lightest supersymmetric particle,
for example the neutralino, cannot be produced thermally, otherwise the
universe will be over-closed. So we have to require that the reheating
temperature be below $10^{10}$~GeV. However, there is the interesting
possibility that the LSP neutralino is produced non-thermally in suitable
amount, and may still be a dark matter candidate.  Then neutralino
annihilation may have cosmological consequences.

If $R$-parity is violated, tiny neutrino masses can be generated naturally
at one-loop.  However, for naturalness, we have to suppress the
tree-level contributions to the neutrino masses from the $\mu_i L_i H_u$ terms
in the superpotential, where $\mu_i$ must be smaller than
10~GeV~\cite{Grossman:2003gq}.  The neutrino masses and bilarge mixings deserve
further detailed studies.

\section{ The Other Axion Models with High-Scale\\ Supersymmetry Breaking}

We now briefly discuss the other axion models with high-scale
supersymmetry breaking, where the axion can be the dark matter
candidate and the axion solution to the strong CP problem may be
stabilized similarly to the discussions in Section~\ref{sec:stab}.
However, the Peccei-Quinn symmetry can be broken naturally
at about $10^{11}$ GeV if and only if the supersymmetry breaking scale
is intermediate.

\subsection{DFSZ Axion Model with Intermediate-Scale Supersymmetry Breaking}

Since we consider approximate universal supersymmetry breaking, {\it i.e.}, the
mass differences among the squarks, sleptons, gauginos, Higgsinos and the
$A$-terms are within one order of magnitude (within one-loop
suppressions), we must introduce additional particles to achieve gauge
coupling unification. Similar to the discussions in Section III, we can
introduce two pairs of SM vector-like particles with SM quantum numbers
((${\bf 3}, {\bf 2}, {\bf 1/6}$), (${\bf \overline{3}}$, ${\bf 2}$, ${\bf
  -1/6}$)) and ((${\bf 3}$, ${\bf 1}$, ${\bf -1/3}$), (${\bf \overline{3}}$,
${\bf 1}$, ${\bf 1/3}$)).  Alternatively, we can introduce two chiral
multiplets $\Phi_8$ and $\Phi_3$ with quantum numbers (${\bf 8}, {\bf 1}, {\bf
  0}$) and (${\bf 1}, {\bf 3}, {\bf 0}$), and one pair of the SM vector-like
fields with quantum numbers ((${\bf 1}$, ${\bf 2}$, ${\bf -1/2}$), (${\bf 1}$,
${\bf 2}$, ${\bf 1/2}$)).  The masses for these extra particles are around the
intermediate scale $10^{11}$ GeV, and gauge coupling unification can be
achieved around $2.0\times 10^{16}$ GeV.

The stabilization of the axion solution is similar to that in Section III for
the KSVZ axion model by introducing a discrete $Z_N$ Peccei-Quinn symmetry and
embedding it into the anomalous $U(1)_A$ gauge symmetry, whose anomalies are
cancelled by the Green--Schwarz mechanism.  As in Section III, in order to
generate the quartic coupling for the singlet $S$, we need to couple $S$ to the
extra particles. For example, the superpotential for the additional particles
in the second set of fields is
\begin{eqnarray}
W &=& y_{\Phi_8} S  \Phi_8 \Phi_8 +
 y_{\Phi_3} S \Phi_3 \Phi_3 ~,~\,
\end{eqnarray}
where the Yukawa couplings $y_{\Phi_8}$ and $y_{\Phi_3}$ are of order 1.  Here
$\Phi_8$ and $\Phi_3$ are charged under the $Z_N$ or $U(1)_A$ symmetry.

We emphasize that in this model the intermediate supersymmetry breaking scale
is still directly related to the Peccei--Quinn symmetry breaking scale.

\subsection{ Axion Models with String-Scale Supersymmetry Breaking}

The axion models with string-scale supersymmetry breaking are the Standard
Model plus the KSVZ or DFSZ axion. There are many ways to achieve gauge
coupling unification by introducing various sets of particles with masses from
the TeV scale to the GUT scale.  For simplicity, we only consider the axion model
with TeV-scale extra fermions whose masses can be protected by the chiral
symmetry.  We can introduce two pairs of SM vector-like fermions with quantum
numbers ((${\bf 3}, {\bf 2}, {\bf 1/6}$), (${\bf \overline{3}}$, ${\bf 2}$,
${\bf -1/6}$)) and ((${\bf 3}$, ${\bf 1}$, ${\bf -1/3}$), (${\bf
  \overline{3}}$, ${\bf 1}$, ${\bf 1/3}$)), similar to the extra particle
content in Ref.~\cite{Wagner}. 
  Gauge coupling unification can be achieved around
$2\times 10^{16}$ GeV~\cite{Wagner}.  For the KSVZ axion model with the
second set of fields, we need to introduce SM vector-like particles which form
complete $SU(5)$ multiplets and couple to the SM singlet $S$
so that we can have the QCD anomalous Peccei--Quinn
symmetry and preserve gauge coupling unification.

The stabilization of the axion solution is similar to that in Section III.
 However, we need extra fine-tuning to keep the VEV of $S$ around $10^{11}$ GeV.

\subsection{ Axion Models with Split Supersymmetry }

For axion models with split supersymmetry, the stabilization of the axion
solution and the Peccei-Quinn symmetry breaking
are similar to those in Section III if the supersymmetry breaking scale
is intermediate, {\it i.e.}, from $10^{10}$ GeV to $10^{12}$ GeV. 
However, for axion models where the
supersymmetry breaking scale is higher than $10^{12}$ GeV or lower than
$10^{10}$ GeV, the stabilization of the axion
solution is similar to that in Section~III,
while how to naturally break the  Peccei-Quinn symmetry
 deserves further study.

For the KSVZ axion model with intermediate-scale supersymmetry breaking, we can
introduce one pair of vector-like fields with $SU(5)$ quantum numbers ${\bf
  5}$ and ${\bf \overline{5}}$, respectively, that can couple to the singlet
$S$ via the superpotential $S {\bf 5} {\bf \overline{5}}$.  As a result, gauge
coupling unification is still preserved, and the stabilization of the axion
solution and the Peccei-Quinn symmetry breaking
are the same as those in Section III.  For the DFSZ axion model with
intermediate-scale supersymmetry breaking, we still need to introduce some
adjoint or SM vector-like particles which form complete $SU(5)$ multiplets that
couple to the singlet $S$ to generate its quartic coupling at one-loop.

There are again two possibilities for the axion models with split supersymmetry:
(1)~$R$-parity is violated because the LSP neutralino need not be a dark
matter candidate at all; (2)~$R$-parity is preserved, and both the LSP
neutralino and the axion or dominantly one of them contributes to the dark
matter density.

\section{Conclusions}
\label{sec:conclu}

There exists the possibility of high-scale supersymmetry breaking in string
compactifications where the cosmological constant problem and 
the gauge hierarchy problem can be solved. However, 
the strong CP problem may still be a serious complication for naturalness
in the string landscape. Motivated by these considerations, 
we constructed a supersymmetric KSVZ axion model where the supersymmetry 
breaking scale and the Peccei--Quinn symmetry breaking scale were both around 
$10^{11}$ GeV and the axion could be a viable cold dark matter candidate.  
We considered the discrete $Z_N$ Peccei-Quinn symmetry that could be embedded
into an anomalous $U(1)_A$ gauge symmetry in string constructions where the
anomalies were cancelled by the Green--Schwarz mechanism. This $Z_N$ discrete
symmetry could not be violated by quantum gravitational corrections.  We found $N
\geq ~ 10$ is necessary to ensure that the contributions to the $\overline{\theta}$ term
from the non-renormalizable operators are under control.  We also showed that a
reasonable quartic coupling for the singlet $S$ could be generated from the
one-loop corrections of extra vector-like particles.  Thus, the intermediate
supersymmetry breaking scale was directly connected to the Peccei-Quinn symmetry
breaking scale. In addition, using two-loop renormalization group equation runnings
for the gauge couplings and one-loop renormalization group equation runnings
for the Yukawa couplings, we showed that gauge coupling unification was
achieved at about $2.7\times 10^{16}$ GeV and that the Higgs mass was in
the range 130 GeV to 160 GeV.  We also calculated the Yukawa couplings for
the third family of the SM fermions at the GUT scale and commented on
the possibility of Yukawa coupling unification.  In our model, due to
intermediate-scale supersymmetry breaking, we did not have the dimension-5
operator induced proton decay problem, the moduli problem, and the problematic
supersymmetry flavor and CP violations, etc.  We also pointed out that if the
$R$-parity was an exact symmetry, the LSP neutralino could be produced only
non-thermally and might have cosmological consequences.  If the $R$-parity was
broken, the tiny masses and bilarge mixings for the left-handed neutrinos could 
be realized naturally at one-loop, although we had to suppress the tree-level
contributions to the neutrino masses from the $\mu_i L_i H_u$ terms in the
superpotential.

Furthermore, we briefly discussed the other axion models: the DFSZ axion model
with intermediate-scale supersymmetry breaking where the stabilization of the
axion solution and the Peccei-Quinn symmetry breaking
were similar, the axion models with string-scale supersymmetry
breaking, and the axion models with split supersymmetry.

\begin{acknowledgments}

T. Li would like to thank I. Gogoladze for helpful discussions.  This
research was supported in part by the U.S.~Department of Energy, High
Energy Physics Division, under Contract DE-FG02-95ER40896,
W-31-109-ENG-38 and DE-FG02-96ER40969, in part by the National Science
Foundation under Grant No.~PHY-0070928, and in part by the University
of Wisconsin Research Committee with funds granted by the Wisconsin
Alumni Research Foundation.

\end{acknowledgments}

\appendix
\section{ Renormalization Group Equations}
\label{apdx}
In this Appendix, we give the renormalization group equations
in the SM, and in our supersymmetric KSVZ axion model.
The general formulae for the renormalization group equations
in the SM are given in Refs.~\cite{mac,Cvetic:1998uw}, and 
these for the supersymmetric models
are given in Refs.~\cite{Barger:1992ac,Barger:1993gh,Martin:1993zk}.

First, we summarize the  renormalization group equations
in the SM.
The two-loop renormalization group equations for the gauge couplings are
\begin{eqnarray}
(4\pi)^2\frac{d}{dt}~ g_i=g_i^3b_i &+&\frac{g_i^3}{(4\pi)^2}
\left[ \sum_{j=1}^3B_{ij}g_j^2-\sum_{\alpha=u,d,e}d_i^\alpha
{\rm Tr}\left( h^{\alpha \dagger}h^{\alpha}\right) \right] ~,~\,
\label{SMgauge}
\end{eqnarray}
where $t=\ln  \mu$ and $ \mu$ is the renormalization scale.
The $g_1$, $g_2$ and $g_3$ are the gauge couplings
for $U(1)_Y$, $SU(2)_L$ and $SU(3)_C$, respectively,
where we use the $SU(5)$ normalization $g_1^2 \equiv (5/3)g_Y^{ 2}$.
The beta-function coefficients are  
\begin{eqnarray}
&&b=\left(\frac{41}{10},-\frac{19}{6},-7\right) ~,~
B=\pmatrix{\frac{199}{50}&
\frac{27}{10}&\frac{44}{5}\cr \frac{9}{10} & \frac{35}{6}&12 \cr
\frac{11}{10}&\frac{9}{2}&-26} ~,~\\
&&d^u=\left(\frac{17}{10},\frac{3}{2},2\right) ~,~
d^d=\left(\frac{1}{2},\frac{3}{2},2\right) ~,~
d^e=\left(\frac{3}{2},\frac{1}{2},0\right) ~.~\,
\end{eqnarray}

Since the contributions in Eq.~(\ref{SMgauge}) from
the Yukawa couplings arise from  the 
two-loop diagrams, we only need Yukawa coupling
 evolution at the one-loop order.
The one-loop renormalization group equations for Yukawa couplings are
\begin{eqnarray}
(4\pi)^2\frac{d}{dt}~h^u&=&h^u\left( -\sum_{i=1}^3c_i^ug_i^2
+\frac{3}{2}
h^{u \dagger}h^{u}
-\frac{3}{2}
h^{d \dagger}h^{d}
+\Delta_2\right) ~,~\\
(4\pi)^2\frac{d}{dt}~h^d&=&h^d\left( -\sum_{i=1}^3c_i^dg_i^2
-\frac{3}{2}
h^{u \dagger}h^{u}
+\frac{3}{2}
h^{d \dagger}h^{d}
+\Delta_2 \right)~,~\\
(4\pi)^2\frac{d}{dt}~h^e&=&h^e\left( -\sum_{i=1}^3c_i^eg_i^2
+\frac{3}{2}
h^{e \dagger}h^{e}
+ \Delta_2 \right) ~,~\,
\label{SMY}
\end{eqnarray}
where 
\begin{eqnarray}
c^u=\left( \frac{17}{20}, \frac{9}{4}, 8\right) ~,~
c^d=\left( \frac{1}{4}, \frac{9}{4}, 8\right) ~,~
c^e=\left( \frac{9}{4}, \frac{9}{4}, 0\right) ~,~
\end{eqnarray}
\begin{eqnarray}
\Delta_2 &=& {\rm Tr} ( 3h^{u \dagger}h^{u}+3 h^{d \dagger}h^{d}+
h^{e \dagger}h^{e})  ~.~
\end{eqnarray}

The one-loop renormalization group equation for the Higgs quartic coupling is 
\begin{eqnarray}
(4\pi)^2\frac{d}{dt}~\lambda  &=&12 \lambda^2
-\left({9\over 5} g_1^2 + 9 g_2^2 \right) \lambda
+{9\over 4} \left( {3\over {25}} g_1^4 
+ {2\over 5} g_1^2g_2^2 + g_2^4 \right)
+4\Delta_2 \lambda - 4 \Delta_4 ~,~\,
\end{eqnarray}
where
\begin{eqnarray}
\Delta_4 &=& {\rm Tr} \left[ 3 (h^{u \dagger}h^{u})^2+3 (h^{d \dagger}h^{d})^2
+ (h^{e \dagger}h^{e})^2\right]  ~.~
\end{eqnarray}

Second, we summarize the renormalization group equations in 
our supersymmetric KSVZ axion model.
The two-loop renormalization group equations for the gauge couplings are
\begin{eqnarray}
(4\pi)^2\frac{d}{dt}~ g_i &=& g_i^3 (b_i + \Delta b_i)
 +\frac{g_i^3}{(4\pi)^2}
\left[~ \sum_{j=1}^3(B_{ij} + \Delta B_{ij}) g_j^2-\sum_{\alpha=u,d,e}d_i^\alpha
{\rm Tr}\left( y^{\alpha \dagger}y^{\alpha}\right) 
\right.\nonumber\\ &&\left.
{}-d_i^{Q_X} y_{Q_X}^{\dagger} y_{Q_X}  
-d_i^{D_X} y_{D_X}^{\dagger} y_{D_X}
\right] ~,~\,
\label{SUSYgauge}
\end{eqnarray}
where $\Delta b_i$ and $\Delta B_{ij}$ are the contributions
from the extra two pairs of the SM vector-like particles.
The beta-function coefficients are 
\begin{eqnarray}
&&b=\left(\frac{33}{5},1,-3\right) ~,~
 \Delta b = \left(\frac{3}{5}, 3, 3\right)~,~\\
&& B=\pmatrix{\frac{199}{25}&
\frac{27}{5}&\frac{88}{5}\cr \frac{9}{5} & 25&24 \cr
\frac{11}{5}&9&14} ~,~
\Delta B=\pmatrix{\frac{3}{25}&
\frac{3}{5}&\frac{16}{5}\cr \frac{1}{5} & 21 & 16 \cr
\frac{2}{5} & 6 & 34} ~,~\\
&&d^u=\left(\frac{26}{5},6,4\right) ~,~
d^d=\left(\frac{14}{5},6,4\right) ~,~
d^e=\left(\frac{18}{5},2,0\right) ~,~ \\
&& d^{Q_X} =\left(\frac{2}{5},6,4\right) ~,~
d^{D_X} =\left(\frac{4}{5},0,2\right)~.~
\end{eqnarray}

The one-loop renormalization group equations for Yukawa couplings are
\begin{eqnarray}
(4\pi)^2\frac{d}{dt}~y^u&=& y^u
\left[ 3 y^{u \dagger} y^{u}+ y^{d \dagger} y^{d}
+3{\rm Tr}( y^{u \dagger} y^{u}) 
-\sum_{i=1}^3c_i^ug_i^2 \right]~,~\\
(4\pi)^2\frac{d}{dt}~y^d&=& y^d
\left[ y^{u \dagger} y^{u} + 3 y^{d \dagger} y^{d}
+{\rm Tr}(3 y^{d \dagger} y^{d}
+ y^{e \dagger} y^{e}) 
-\sum_{i=1}^3c_i^dg_i^2 \right]~,~\\
(4\pi)^2\frac{d}{dt}~y^e&=& y^e
\left[ 3 y^{e \dagger} y^{e}+{\rm Tr}(3 y^{d \dagger} y^{d}
+ y^{e \dagger} y^{e}) 
-\sum_{i=1}^3c_i^eg_i^2 \right] ~,~\\
(4\pi)^2\frac{d}{dt}~y_{Q_X} &=& y_{Q_X}
\left[  8 y_{Q_X}^{\dagger} y_{Q_X}
+ 3 y_{D_X}^{\dagger} y_{D_X}
-\sum_{i=1}^3 c_i^{Q_X} g_i^2 \right] ~,~\\
(4\pi)^2\frac{d}{dt}~y_{D_X} &=& y_{D_X}
\left[  6 y_{Q_X}^{\dagger} y_{Q_X}
+ 5 y_{D_X}^{\dagger} y_{D_X}
-\sum_{i=1}^3 c_i^{D_X} g_i^2 \right] ~,~
\end{eqnarray}
where
\begin{eqnarray}
&& c^u=\left( \frac{13}{15}, 3, \frac{16}{3}\right) ~,~
c^d=\left( \frac{7}{15}, 3, \frac{16}{3}\right) ~,~
c^e=\left( \frac{9}{5}, 3, 0\right) ~,~ \\
&& c^{Q_X} =\left( \frac{1}{15}, 3, \frac{16}{3}\right)~,~
c^{D_X} =\left( \frac{4}{15}, 0, \frac{16}{3}\right)~.~\,
\end{eqnarray}

\end{document}